\documentstyle[epsf,aps,psfig]{revtex}
\renewcommand{\thefootnote}{\fnsymbol{footnote}}
\begin{document}
\begin{flushright}

Duke preprint DU--TH--143
\end{flushright}
\vspace*{1cm} 
\setcounter{footnote}{1}
\begin{center}
{\Large\bf Gluon Radiation and Coherent States in Ultrarelativistic
Nuclear Collisions}
\\[1cm]
Sergei G.\ Matinyan$^{1,2}$, Berndt M\"uller$^1$, and Dirk H.\ Rischke$^1$ 
\\ ~~ \\
{\it $^1$ Department of Physics,  Duke University} \\ 
{\it Box 90305, Durham, North Carolina 27708-0305} \\ ~~ \\
{\it $^2$ Yerevan Physics Institute, Yerevan 375036, Armenia} \\ ~~ \\ ~~ \\
\end{center}
\begin{abstract} 
We explore the correspondence between classical gluon radiation and
quantum radiation in a coherent state for gluons produced in 
ultrarelativistic nuclear collisions. The expectation value of
the invariant momentum distribution of gluons in the 
coherent state is found to agree with the gluon number distribution 
obtained classically from the solution of the Yang--Mills equations. 
A criterion for the applicability of the coherent state formalism
to the problem of radiation in ultrarelativistic nucleus--nucleus
collisions is discussed. This criterion is found to be fulfilled for 
midrapidity gluons with perturbative transverse momenta larger than 
$\sim 1-2$ GeV and produced in collisions between valence partons.

\ \\ PACS number(s): 12.38.Bx, 12.38.Aw, 24.85.+p, 25.75-q

\end{abstract}
\renewcommand{\thefootnote}{\arabic{footnote}}
\setcounter{footnote}{0}

\section{Introduction}

The planned experiments at Brookhaven and CERN
aimed at producing a dense quark--gluon plasma \cite{muller}
in highly energetic nucleus--nucleus collisions \cite{harrismuller}
have generated interest in theoretical descriptions of the formation 
of such a state. One
approach, the parton cascade model \cite{klaus}, describes the
equilibration process as a sequence of binary collisions of
perturbative quanta in quantum chromodynamics (QCD).  An unresolved
issue of this model is how the decoherence of the initial
nuclear parton distributions occurs.  Two mechanisms have been
proposed: the formation of a large number of mini-jets in the nuclear
collision \cite{Kajantie,HIJING,Eskola} and the radiation of soft 
gluons from the colliding valence quarks \cite{Alex}. 
We are here concerned with the second mechanism. 
We note that it was recently argued in \cite{MGLMcL} that 
the two approaches may be intimately related and lead to almost identical 
predictions in certain limits.

The gluon radiation mechanism was originally \cite{Alex} formulated in the
framework of the classical approach proposed by
McLerran and Venugopalan \cite{Larry}.  In their model, the small--$x$ 
gluon distribution of a large nucleus is described as arising
from the classical gauge field created by a random distribution
of color sources representing the valence quarks of the nucleons
and moving along the light-cone, and by the quantum fluctuations 
around this field.
Kovchegov \cite{yuri} has shown how the classical field can be
obtained from a model of the fast moving nucleus as a collection of
color dipoles and has identified the limitations of the classical
approximation \cite{yuri'}.

When two nuclei collide, a part of their classical fields is scattered
on-shell, describing the emission of real gluons
\cite{Alex}.  A detailed derivation of this process has recently been
given to lowest and next-to-lowest order in the QCD coupling constant 
$g$ \cite{yuridirk}. Using semi-classical arguments, the radiated field 
energy can be related to the gluon multiplicity
distribution as function of rapidity and transverse momentum.  The
result was shown to agree with the prediction of the perturbative,
lowest order quantum calculation of soft gluon emission by 
colliding color charges \cite{gunion}.

In this work we demonstrate that classical gluon radiation as
computed in \cite{Alex,yuridirk} is closely related to
quantum radiation of gluons in a coherent state, and in fact
gives the same result for the gluon number distribution. 
In the coherent state, gluons are emitted in independent 
binary interactions between the colliding valence quarks of the two nuclei.
Thus, if the number of emitted gluons is only a
small fraction of the number of colliding color charges, 
correlations in the emission of individual gluons, such as arising
from multiple collisions of the same charge, are
negligible and the coherent state approach agrees with a quantum
calculation of the multiple scattering process.  
If, on the other hand, a significant fraction of the colliding charges 
radiates, such correlations could become important and 
corrections to the coherent state calculation have to be accounted for. 
On the basis of this argument we estimate that for central 
$Au+Au$--collisions at RHIC energies the coherent state approach is 
applicable to describe gluon radiation, if the gluons are produced in
collisions between valence partons and at midrapidity with 
perturbative transverse momenta above about $1-2$ GeV. 

The remainder of this paper is organized as follows. In Section II we
briefly outline the results of \cite{yuridirk} as far as they
are of relevance to this work. Section III consists of two parts.
In the first, the coherent state corresponding to gluon radiation
from collisions of classical color charges is discussed.
The second part interprets the results in terms of quantum diagrams.
Conditions for the validity of the coherent state approach are discussed.
Section IV contains the derivation of the event-averaged 
gluon number distribution in the coherent state formalism. The result
is found to agree with that of Ref.\ \cite{yuridirk}. We then generalize
it to nuclear collisions at finite impact parameter.
Section V concludes this work with a numerical evaluation of the gluon
number distribution and a quantitative discussion of 
the validity of the coherent state approach for the description
of gluon radiation in ultrarelativistic nuclear collisions.

Our units are $\hbar = c =1$, and the metric tensor is $g^{\mu \nu} =
{\rm diag} (+,-,-,-)$. Light-cone coordinates are defined as
$a_\pm \equiv (a^0 \pm a^z)/\sqrt{2},\, \partial_\mp \equiv
\partial / \partial x_\pm$. The notation for 3--vectors is
${\bf a} = (a^x,a^y,a^z)$ and for transverse vectors
$\underline{a} = (a^x,a^y)$.

\section{Classical gluon radiation in ultrarelativistic nuclear collisions}

\begin{figure}
\begin{center}
\epsfxsize=7cm
\epsfysize=7cm
\leavevmode
\hbox{ \epsffile{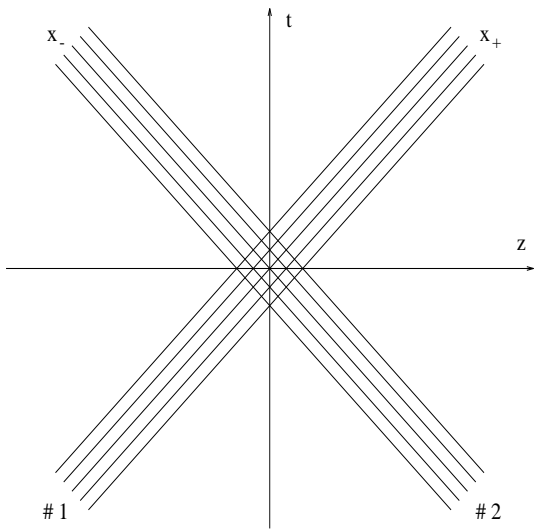}}
\end{center}
\caption{The nuclear collision as envisaged here.}
\label{coll}
\end{figure}

In this section, we briefly summarize the results of \cite{yuridirk},
as far as they are required for the following.
In covariant (Lorentz) gauge, $\partial_\mu A^\mu = 0$,
the classical Yang--Mills equations,
\begin{equation} \label{eom}
D_\mu F^{\mu \nu} = J^{\nu}\,\, ,
\end{equation}
where $D_\mu  \equiv  \partial_\mu -ig\, [\, A_\mu \, , \, \cdot \, ] 
\, , \,\, F^{\mu \nu}  \equiv   \partial^\mu A^\nu - \partial^\nu A^\mu - 
ig\, [\, A^\mu\, , \, A^\nu \, ] $,
and $J^\nu$ is a classical source current, can be cast into the form
\begin{equation} \label{eom2}
\Box A^{\mu} = J^\mu + ig \, [\, A_\nu \, , \, \partial^\nu A^\mu + 
F^{\nu \mu} \, ] \,\, ,
\end{equation}
where $\Box$ is the d'Alembertian operator. In this form, the equations
can be readily solved perturbatively order by order in the
strong coupling constant, $g$. 
We expand
\begin{equation} \label{expand}
A_\mu = \sum_{k=0}^\infty \,  A^{(2k+1)}_\mu \,\,\,\, , \,\,\,\,\,
J_\mu = \sum_{k=0}^\infty \,  J^{(2k+1)}_\mu \,\,\, ,
\end{equation}
where $A^{(n)}_\mu, \, J^{(n)}_\mu$ are the contributions of order $g^n$ 
to the gluon field and source current, respectively.
[Eq.\ (\ref{expand}) takes into account that
only odd integers of $g$ occur in this expansion.]
Then, the lowest and next-to-lowest order solutions obey
\begin{mathletters} \label{eom3}
\begin{eqnarray}
\Box A_{\mu}^{(1)} & = & J_\mu^{(1)} \equiv \tilde{J}_\mu^{(1)} \,\,\, , \\
\Box A_{\mu}^{(3)} & = & J_\mu^{(3)} + ig \, [ \, A^{(1)\nu} \, , \,
\partial_\nu A_\mu^{(1)} + F_{\nu \mu}^{(1)} \, ] 
\equiv \tilde{J}_\mu^{(3)} \,\,\, . \label{eom3b}
\end{eqnarray}
\end{mathletters}
These equations are linear to each successive order in $g$ and can therefore
be solved with the method of Green functions:
\begin{equation} \label{clsol}
A_{\mu}^{(2k+1)} (x) = \int d^4 x'\, G_r (x - x')\, \tilde{J}_{\mu}^{(2k+1)} 
(x')\,\, , \,\,\,\, k=0,1 \,\, .
\end{equation}
The retarded Green function reads in coordinate and
momentum space \cite{itzykson}:
\begin{equation} \label{Green}
G_r(x) = \frac{1}{2 \pi} \, \theta(t)\, \delta(x^2)\,\,\,\, , \,\,\,\,\,
\tilde{G}_r(k) = - \,\frac{1}{k^2 + i \epsilon k_0 }\,\, .
\end{equation}
Let us now consider a collision of two nuclei with mass numbers 
$A_1, \, A_2$, moving towards each other with ultrarelativistic velocities, 
$v_{1,2} \simeq \pm 1$, along the $z$--axis.  The nuclei are
taken as ensembles of nucleons \cite{yuri,yuridirk}, cf.\
Fig.\ \ref{coll}.

In order to simplify the color algebra we assume that
each ``nucleon'' consists of a quark--antiquark
pair. These valence quarks and antiquarks are confined inside the
nucleons (visualized as spheres of equal radius in the rest frame of
each nucleus).  In order to construct the solution, nucleons inside
the nucleus and valence charges inside the nucleons are assumed to be
``frozen'', i.e., they have definite light-cone (and transverse)
coordinates. We assume the charges to move along recoilless 
trajectories, therefore, their coordinates will not change throughout 
the calculation. We
label the coordinates of the quarks in nucleus 1 by $x_{-i},\,
\underline{x}_i$, $i=1, \ldots ,A_1$, and those of nucleus 2 by
$y_{+j},\, \underline{y}_j$, $j=1, \ldots ,A_2$.  Antiquark
coordinates follow this notation with an additional prime. 

Then, the lowest order classical current is
a sum of the currents for each individual nucleus, as given in 
Eq.\ (3) of Ref.\ \cite{yuridirk}.
The lowest order solution, $A_\mu^{(1)}$, and the associated
field strength tensor, $F_{\mu \nu}^{(1)}$, were given in
Ref.\ \cite{yuridirk}, Eqs.\ (4,5). This solution is identical to
the one of the corresponding Abelian problem, i.e., the nuclei
pass through each other without interacting. The fields generated
by the valence charges simply superpose and no gluons are radiated.

With this lowest order solution and the classical current to next-to-lowest
order in $g$, $J^{(3)}_\mu$, one can compute the next-to-lowest order 
solution, $A_\mu^{(3)}$.
The classical current $J^{(3)}_\mu$ was obtained in \cite{yuridirk}
from covariant current conservation, $D_\mu\, J^\mu = 0$, and the 
assumption of recoilless trajectories, see Eq.\ (8) of \cite{yuridirk}.
The right-hand side of Eq.\ (\ref{eom3b}) then reads in coordinate space
\begin{equation} \label{current03x}
\tilde{J}^{(3)a}_{\mu}(x) = \sum_{i,j} \, \left[ \tilde{J}^{(3)a}_\mu
(x;x_i,y_j) - \tilde{J}^{(3)a}_\mu (x;x_i',y_j) -\tilde{J}^{(3)a}_\mu
(x;x_i,y_j') + \tilde{J}^{(3)a}_\mu(x;x_i',y_j') \right] \,\, ,
\end{equation}
where
\begin{mathletters}\label{currect3x}
\begin{eqnarray}
\tilde{J}_+^{(3)a}(x;x_i,y_j) 
 & = & {g^3 \over { (2 \pi)^2 }} \, f^{abc}\, (T_i^b)
 \, (\tilde{T}_j^c) \left( - \,\, 2\pi\, \ln (|{\underline x}_i -
 {\underline y}_j| \lambda) \, \delta (x_- - x_{-i})\, \theta (x_+ -
 y_{+j}) \, \delta({\underline x}-{\underline x}_i) \right.  \\ & &
 \hspace*{3.2cm} \left. + \,\, \ln (|{\underline x} - {\underline
 x}_i| \lambda) \, \ln (|{\underline x} - {\underline y}_j| \lambda)
 \, \partial_+ \delta(x_- - x_{-i})\, \delta (x_+ - y_{+j})
 \right)\,\, , \nonumber \\
\tilde{J}_-^{(3)a}(x;x_i,y_j) 
& = & {g^3 \over { (2 \pi)^2 }} \, f^{abc}\, (T_i^b)
 \, (\tilde{T}_j^c) \left( \,\, 2\pi\, \ln (|{\underline x}_i -
 {\underline y}_j| \lambda) \, \theta (x_- - x_{-i})\, \delta (x_+ -
 y_{+j}) \, \delta({\underline x}-{\underline y}_j) \right.  \\ & &
 \hspace*{3.2cm}\left. - \,\, \ln (|{\underline x} - {\underline x}_i|
 \lambda) \, \ln (|{\underline x} - {\underline y}_j| \lambda) \,
 \delta(x_- - x_{-i})\, \partial_- \delta (x_+ - y_{+j}) \right) \,\,
 , \nonumber \\
\underline{\tilde{J}}^{(3)a}(x;x_i,y_j) 
& = & {g^3 \over { (2 \pi)^2 }}\,
 f^{abc}\, (T_i^b) \, (\tilde{T}_j^c) \, \delta(x_- - x_{-i}) \,
 \delta(x_+-y_{+j}) \\ & & \hspace*{3.2cm} \times \,\, \left(\, \ln(
 |{\underline x}- {\underline y}_j| \lambda) \, \frac{{\underline x}-
 {\underline x}_i}{|{\underline x}- {\underline x}_i|^2} - \ln(
 |{\underline x}- {\underline x}_i| \lambda) \, \frac{{\underline x}-
 {\underline y}_j}{|{\underline x}- {\underline y}_j|^2} \right) \,\,
 . \nonumber
\end{eqnarray}
\end{mathletters}
Here, $(T_i^a),\, (\tilde{T}_j^b)$ are color matrices which
represent the color charge of the quarks in the
color space of nucleon $i$ of nucleus 1 and nucleon $j$ of
nucleus 2, while $f^{abc}$ are the structure constants of $SU(N_c)$.
The antiquarks have the opposite color charge, $-(T_i^a),\,
-(\tilde{T}_j^b)$, which explains the relative signs in Eq.\
(\ref{current03x}), and ensures color neutrality of each nucleon. 
$\lambda$ is an infrared cut-off and acts as gauge parameter for
the lowest order solution.

The next-to-lowest order solution is then computed as
\begin{equation} \label{clsol2}
A_{\mu}^{(3)a} (x) = - \int \frac{ d^4 k}{(2 \pi)^4}\, \frac{e^{- i k
 \cdot x} }{k^2 + i \epsilon k_0 } \, \tilde{J}^{(3)a}_{\mu}(k)\,\, ,
\end{equation}
where 
\begin{equation} \label{current03k}
\tilde{J}^{(3)a}_{\mu}(k) = \sum_{i,j} \, \left[ \tilde{J}^{(3)a}_\mu
(k;x_i,y_j) - \tilde{J}^{(3)a}_\mu (k;x_i',y_j) -\tilde{J}^{(3)a}_\mu
(k;x_i,y_j') + \tilde{J}^{(3)a}_\mu(k;x_i',y_j') \right] \,\, ,
\end{equation}
and
\begin{mathletters} \label{current3k}
\begin{eqnarray} \label{current3ka}
\tilde{J}_{+}^{(3)a} (k;x_i,y_j) & = & \frac{g^3}{ (2 \pi)^2}  f^{abc}
 (T_i^b) \, (\tilde{T}_j^c)\, e^{i (k_+ x_{-i} + k_- y_{+j} -
 \underline{k} \cdot \underline{y}_j)} \! \int \!\! d^2 \underline{q} \,
 e^{-i \underline{q} \cdot (\underline{x}_i - \underline{y}_j)} \,
 \frac{1}{(\underline{k} - \underline{q})^2}  \left[ \frac{ i }{k_-
 + i \epsilon} - \frac{ i k_+ }{\underline{q}^2} \right] 
 ,\! \\ \label{current3kb}
 \tilde{J}_{-}^{(3)a} (k;x_i,y_j) & = & \frac{g^3}{ (2 \pi)^2} f^{abc}
 (T_i^b) \, (\tilde{T}_j^c)\, e^{i (k_+ x_{-i} + k_- y_{+j} -
 \underline{k} \cdot \underline{y}_j)}\! \int\!\! d^2  \underline{q} \,
 e^{-i \underline{q} \cdot (\underline{x}_i - \underline{y}_j)} \,
 \frac{1}{\underline{q}^2}  \left[ \frac{- i }{k_+ + i \epsilon} +
 \frac{i k_- }{ (\underline{k} - \underline{q})^2} \right] 
 ,\! \\
 \underline{\tilde{J}}^{(3)a} (k;x_i,y_j) & = & \frac{g^3}{ (2 \pi)^2}
 f^{abc}(T_i^b) \, (\tilde{T}_j^c)\, e^{i (k_+ x_{-i} + k_- y_{+j}
 - \underline{k} \cdot \underline{y}_j)} \!\int \!\! d^2 \underline{q} \,
 e^{-i \underline{q} \cdot (\underline{x}_i - \underline{y}_j)} \,
 \frac{i (2\underline{q} - \underline{k})}{ \underline{q}^2 (
 \underline{k}- \underline{q})^2} \,\,\, . \label{current3kc}
\end{eqnarray} 
\end{mathletters}
Note that an explicit expression for
the solution $A_\mu^{(3)a}(x)$ was given in Ref.\ \cite{yuridirk}, Eqs.\
(21--24). The field (\ref{clsol2})
contains a piece associated with the change of color of a charge
when it collides with the field of another charge.
It arises from the pole $k_\pm = -i\epsilon$ in Eqs.\ 
(\ref{current3ka},\ref{current3kb})
and thus does not correspond to radiated gluons, 
cf.\ also Eq.\ (25) in \cite{yuridirk}.
The radiated gluon field, on the other hand, 
arises from the poles of the retarded
propagator in Eq.\ (\ref{clsol2}), i.e., it corresponds to on-shell
gluons, as one would expect,
\begin{equation} \label{Arad0}
A_{\mu {\rm \,rad}}^{(3)a}(x) = \sum_{i,j} \, \left[
   A_{\mu {\rm \,rad}}^{(3)a}(x;x_i,y_j)
 - A_{\mu {\rm \,rad}}^{(3)a}(x;x_i',y_j)
 - A_{\mu {\rm \,rad}}^{(3)a}(x;x_i,y_j')
 + A_{\mu {\rm \,rad}}^{(3)a}(x;x_i',y_j') \right] \,\, ,
\end{equation}
where
\begin{equation} \label{Arad}
A_{\mu{\rm \,rad}}^{(3)a} (x;x_i,y_j) = \theta(t - t_{ij})\, \int d \tilde{k}
\left[ \, i\, \tilde{J}_{\mu}^{(3)a}(\omega,{\bf k};x_i,y_j) \,\, 
e^{-i \omega t + i{\bf k} \cdot {\bf x}} + {\rm c.c.} \right] \,\, ,
\end{equation}
$t_{ij} \equiv (x_{-i} + y_{+j})/\sqrt{2}$ is the time when the 
collision between the quarks of nucleon $i$ and $j$ happens, 
and $ d\tilde{k} \equiv d^3 {\bf k}/[(2 \pi)^3 2
\omega]$, with $\omega = |{\bf k}| $. Obviously, this
field vanishes prior to the collision.

\section{Gluon radiation as a coherent state}

\subsection{Formalism}

In the preceding section and Refs.\ \cite{Alex,yuridirk}, 
the gluons produced in the collision of the two nuclei were treated as
a classical field which solves the Yang--Mills equations. In contrast,
in this section we want to describe the gluons as quanta radiated
in the presence of a classical source, which we will take to be the
current (\ref{current03x}). The way to solve a problem of this form can
be found in standard textbooks \cite{itzykson}.

Adapting the usual treatment to our case of a nuclear collision, we
note that at $t \rightarrow - \infty$ the current (\ref{current03x}) is
zero and no real gluons are present. The gluonic Fock space ``in''--state  
is therefore the physical vacuum, $|\, 0 \, \rangle$. 
When the two nuclei collide, 
the $c$--number source current (\ref{current03x}) produces real gluons such 
that at $t \rightarrow \infty$, the ``out''--state of the system is the 
{\em coherent\/} state \cite{itzykson,slotta}
\begin{equation} \label{coh}
| \, \tilde{J}\,\rangle \equiv e^{-\bar{N}/2} \, \exp \left[
-i \int d\tilde{k} \sum_{\lambda=1}^2 \sum_{a=1}^8 \,
\tilde{J}^{(3)a}(\omega,{\bf k})\cdot
\epsilon({\bf k},\lambda,a) \, \hat{c}^\dagger ({\bf k},\lambda,a)
\right] \, | \, 0\, \rangle\,\,\, ,
\end{equation}
where
\begin{equation}
\tilde{J}_\mu^{(3)a}(\omega,{\bf k}) = \int d^4x\, e^{i(\omega t -
{\bf k} \cdot {\bf x})}\, \tilde{J}_\mu^{(3)a}(x)
\end{equation}
is the on-shell Fourier transform of the classical current (\ref{current03x}). 
It is identical with the current (\ref{current03k}), taken with 
on-shell momenta $k^\mu =(\omega, {\bf k})$, as implied by our notation. 
The $\epsilon^\mu ({\bf k},\lambda,a)$ are polarization vectors for (real)
gluons with 3--momentum ${\bf k}$, polarization $\lambda$, and
color $a$, obeying
\begin{equation}
k \cdot \epsilon({\bf k},\lambda,a) = 0\,\, , \,\,\,
\epsilon({\bf k},\lambda,a) \cdot \epsilon({\bf k},\lambda',a) =
g^{\lambda \lambda'}\,\, .
\end{equation}
The creation operators $\hat{c}^\dagger ({\bf k},\lambda,a)$ and
the corresponding annihilation operators $\hat{c} ({\bf k},\lambda,a)$
for such gluons fulfill the commutation relation \cite{itzykson}:
\begin{equation}
\left[ \hat{c} ({\bf k},\lambda,a)\, ,\, \hat{c}^\dagger 
({\bf k'},\lambda',b)
\right] = - 2 \omega\, (2\pi)^3\, g^{\lambda \lambda'}\, \delta_{ab} \,
\delta ({\bf k}-{\bf k'})\,\, .
\end{equation}
The first exponential factor in Eq.\ (\ref{coh}) ensures unitarity.
It provides the normalization of the coherent state 
$|\, \tilde{J}\, \rangle$, with
\begin{equation} \label{barn}
\bar{N} =- \int d\tilde{k}\, \left[\tilde{J}^{(3)a}(\omega, {\bf k}) 
\right]^* \cdot \, \tilde{J}^{(3)a}(\omega, {\bf k})
\end{equation}
being the expectation value of the gluon number 
(with all possible colors and polarizations) in the coherent state.
(The minus sign arises from our choice of metric, $\bar{N}$
is, of course, positive definite.)
Since the color quantum number of final states in the collision 
will not be observed, it is not necessary to decompose
the coherent state $|\, \tilde{J} \, \rangle$ in terms of states with
good color quantum number \cite{horn}, i.e., for our purpose
it is sufficient to consider the ``conventional'' coherent state (\ref{coh}).

The coherent state (\ref{coh}) is an eigenstate of
the annihilation operator $\hat{c}({\bf k},\lambda,a)$,
\begin{equation}
\hat{c} ({\bf k},\lambda,a)\, |\, \tilde{J}\,\rangle = 
- i \, \tilde{J}^{(3)a}(\omega,{\bf k}) \cdot
\epsilon({\bf k},\lambda,a) \, |\, \tilde{J}\, \rangle\,\,\, .
\end{equation}
Therefore, the expectation value of the number of gluons with
momentum ${\bf k}$, polarization $\lambda$, and color $a$ in the
coherent state $|\tilde{J} \, \rangle$ is 
\begin{equation}
\langle \, \tilde{J}\, | \, \hat{c}^\dagger ({\bf k},\lambda,a)\,
\hat{c} ({\bf k},\lambda,a)\, | \, \tilde{J}\, \rangle 
= \left[\tilde{J}^{(3)a}(\omega,{\bf k})\right]^* \cdot 
\epsilon({\bf k},\lambda,a)\, \tilde{J}^{(3)a}(\omega,{\bf k}) \cdot 
\epsilon({\bf k},\lambda,a)
\end{equation}
(no summation over $a$).
Summing over colors and (transverse) polarizations, using
\begin{equation}
\sum_{\lambda=1}^2 \epsilon^\mu ({\bf k},\lambda,a)
\, \epsilon^\nu ({\bf k},\lambda,a) =
-g^{\mu \nu} - \frac{k^\mu k^\nu}{(n \cdot k)^2} +
\frac{k^\mu n^\nu + n^\mu k^\nu}{n \cdot k}
\end{equation}
with an arbitrary time-like unit vector $n^\mu$,
and current conservation
\begin{equation}
k \cdot \tilde{J}^{(3)a}(\omega,{\bf k}) = 0
\end{equation}
[this relation can be readily checked with the explicit form 
(\ref{current03k},\ref{current3k}) for the current $\tilde{J}^{(3)a}_\mu$],
we obtain the invariant momentum distribution for
the expectation value of the total gluon number 
in the coherent state $|\, \tilde{J}\, \rangle$,
\begin{equation} \label{dbarndk}
\frac{d\bar{N}}{dy\, d^2 \underline{k}} 
= \frac{1}{2(2\pi)^3}\, \sum_{\lambda=1}^2 \sum_{a=1}^8 \,
\langle \, \tilde{J}\, | \, \hat{c}^\dagger ({\bf k},\lambda,a)\,
\hat{c} ({\bf k},\lambda,a)\, | \, \tilde{J}\, \rangle 
\equiv - \frac{1}{2(2 \pi)^3}\, \left[\tilde{J}^{(3)a}(\omega, {\bf k}) 
\right]^* \cdot \, \tilde{J}^{(3)a}(\omega, {\bf k})\,\, .
\end{equation}
Here $y=\frac{1}{2} \ln [k_+/k_-]$ is the (longitudinal) 
rapidity of the gluons. Again, integrating over invariant momentum
space $dy\, d^2 \underline{k}$, we obtain the expectation value
of the total gluon number $\bar{N}$ in the coherent state
$| \, \tilde{J} \, \rangle$, i.e., Eq.\ (\ref{barn}).

\subsection{Interpretation}

The coherent state $| \, \tilde{J} \, \rangle$, Eq.\ (\ref{coh}), 
is a superposition of 0--, 1--, 2--, $\ldots$ , $n$--, $\ldots$ gluon states,
\begin{eqnarray} \label{coh2}
| \, \tilde{J} \, \rangle & = & e^{-\bar{N}/2}\, \sum_{n=0}^\infty
\frac{(-i)^n}{n!} \prod_{i=1}^n\left[ \int d\tilde{k}_i \,
\sum_{\lambda_i=1}^2 \sum_{a_i=1}^8\, \tilde{J}^{(3)a_i}
(\omega_i, {\bf k}_i) \cdot \epsilon ({\bf k}_i,\lambda_i,a_i)\,
\hat{c}^\dagger ({\bf k}_i,\lambda_i,a_i) \right]\, |\, 0 \, \rangle \\
& = & e^{-\bar{N}/2} \left( |\, 0\, \rangle - i \int d\tilde{k} 
\sum_{\lambda=1}^2 \sum_{a=1}^8\, \tilde{J}^{(3)a}
(\omega, {\bf k}) \cdot \epsilon ({\bf k},\lambda,a)\,\,
|\, {\bf k},\lambda,a \, \rangle \right. \nonumber \\
&  & \left.- \frac{1}{2} \int d\tilde{k}
\, d\tilde{k}'\, \sum_{\lambda,\lambda'=1}^2 \sum_{a,b=1}^8\, 
\tilde{J}^{(3)a} (\omega, {\bf k}) \cdot \epsilon 
({\bf k},\lambda,a)\, \tilde{J}^{(3)b} (\omega', {\bf k}') 
\cdot \epsilon ({\bf k}',\lambda',b)\,\, |\, {\bf k},\lambda,a;\,
{\bf k}',\lambda',b \, \rangle  + \ldots \right)\,\, . \nonumber
\end{eqnarray}
The amplitude for a final state of $n$ gluons with momenta 
${\bf k}_1,\ldots ,\,  {\bf k}_n$, polarizations 
$\lambda_1,\ldots ,\, \lambda_n$, and colors $a_1,\ldots , \, a_n$ 
in the coherent state (\ref{coh2}) is
\begin{eqnarray} \label{ampl}
{\cal M}({\bf k}_1,\lambda_1,a_1; \ldots ;\, {\bf k}_n, \lambda_n, a_n) 
& \equiv &  \langle \,{\bf k}_1,\lambda_1,a_1;\ldots;\, {\bf k}_n, 
\lambda_n, a_n \, | \, \tilde{J} \, \rangle \\
& = & e^{-\bar{N}/2} \, \prod_{i=1}^n \, \left[ -i \, \tilde{J}^{(3)a_i}
(\omega_i, {\bf k}_i) \cdot \epsilon ({\bf k}_i,\lambda_i,a_i) \right] \,\, .
\nonumber
\end{eqnarray}
Thus, the probability to find $n$ gluons in the final state 
{\em regardless\/} of spin, polarization, and color is
\begin{equation}
P_n = \frac{1}{n!} \int  d\tilde{k}_1 \cdots d\tilde{k}_n \, 
\sum_{\lambda_1,\ldots , \lambda_n}\, \sum_{a_1,\ldots , a_n}\,
\left| \, {\cal M}({\bf k}_1,\lambda_1,a_1;\ldots ;\, 
{\bf k}_n, \lambda_n, a_n) \, \right|^2 \,\, ,
\end{equation}
where the prefactor takes into account that the gluons are
indistinguishable Bose particles. A straightforward calculation yields the
well-known result
\begin{equation} \label{Poisson}
P_n = e^{-\bar{N}} \, \frac{\bar{N}^n}{n!}\,\,\, ,
\end{equation}
where $\bar{N}$ is given by Eq.\ (\ref{barn}). The emission of gluons
follows a Poisson probability distribution, or
in other words, it happens in a statistically independent way.
Although classically there are $4\, A_1 A_2$ collisions between the
quarks and antiquarks, and all these collisions act as classical sources for
the gluon field, quantum mechanically there is a certain probability
that no gluon, one gluon, $\ldots$ , $n$, $\ldots$ gluons are emitted in the
nuclear collision. This probability is given by Eq.\ (\ref{Poisson}).

\begin{figure}
\begin{center}
\epsfxsize=15cm
\epsfysize=3cm
\leavevmode
\hbox{ \epsffile{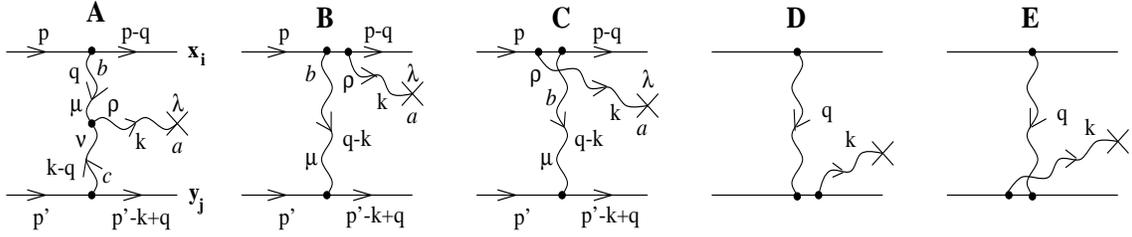}}
\end{center}
\caption{Lowest order diagrams contributing to the amplitude for the 
emission of one gluon.}
\label{diag}
\end{figure}

The amplitude (\ref{ampl}) for a
state where $n$ gluons are emitted during the collision can
also be computed diagrammatically with the usual Feynman rules. Let
us first focus on the case $n=1$.
To lowest order in $g$, the amplitude for the emission of one gluon with 
(on-shell) momentum $k^\mu$, polarization $\lambda$, and color $a$
is given by the diagrams in Fig.\ \ref{diag}. In the recoilless limit
for the quarks, a calculation of these diagrams
with the usual Feynman rules yields the result $-i\, \tilde{J}^{(3)a}
(\omega,{\bf k}) \cdot \epsilon ({\bf k}, \lambda, a)$
[Note that in Ref.\ \cite{yuridirk} the gluon field to order
$g^3$, $A_\mu^{(3)a}(x)$, was calculated from the same set of diagrams. 
The only difference between that calculation and the present one is that 
now the emitted (on-shell) gluon contributes a factor 
$\epsilon^\rho ({\bf k},\lambda,a)$
instead of the (retarded) gluon propagator times a phase
$e^{-ik \cdot x}$ as in \cite{yuridirk},
and there is no integration over the gluon 4--momentum $k^\mu$.]

To lowest order in $g$, the result
for the perturbative amplitude for one--gluon emission is therefore
identical to the amplitude (\ref{ampl}) for $n=1$. [To lowest order,
the prefactor $e^{-\bar{N}/2} \simeq 1$, since $\bar{N} \sim O(g^6)$.]
This is another manifestation of the observation made in
\cite{MGLMcL,yuridirk} that, to order $g^3$, 
classical and quantum calculation give the same result for gluon radiation.
Note, however, that although classically there are $4\, A_1 A_2$ collisions
with gluon emission, the quantum diagram corresponds to {\em only one
single collision\/} between any one charge in nucleus 1 and one in 
nucleus 2 and {\em one single gluon\/} emitted as a result of that
collision. Of course, there are $4\, A_1 A_2$ possible pairs of
colliding charges, corresponding to the $4\, A_1 A_2$ terms in
$\tilde{J}^{(3)a}_\mu$, Eq.\ (\ref{current03k}).
This increased probability for a collision is reflected in the gluon number
distribution which also grows like $4\, A_1 A_2$, cf.\ Section IV.

\begin{figure}
\begin{center}
\epsfxsize=15cm
\epsfysize=4cm
\leavevmode
\hbox{ \epsffile{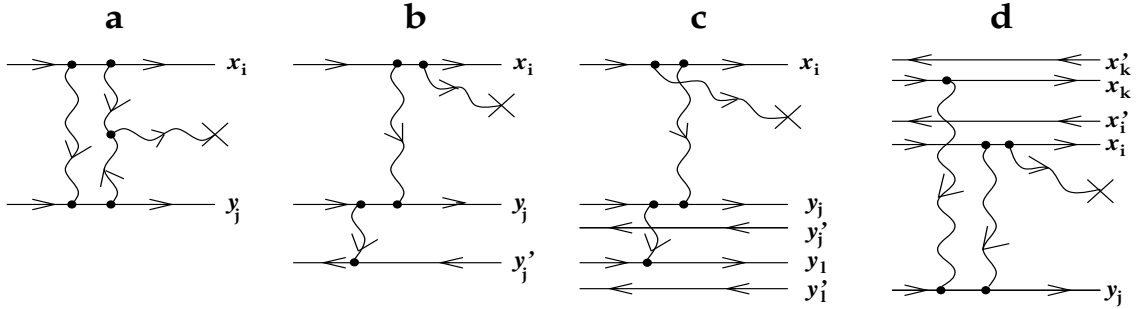}}
\end{center}
\caption{Higher order quantum corrections to the emission process.}
\label{correl}
\end{figure}

To higher order in $g$, there are the following type of quantum corrections
to the one--gluon emission process:
\begin{enumerate}
\item loops as shown in Fig.\ {\ref{correl}}a, 
\item interactions of quark and antiquark inside a nucleon which
participates in the gluon emission process, see for instance Fig.\
{\ref{correl}}b,
\item interactions between quarks and antiquarks of different nucleons
in the same nucleus, cf.\ Fig.\ {\ref{correl}}c, and finally 
\item interactions between quarks and antiquarks
in different nuclei, Fig.\ \ref{correl}d.
\end{enumerate}

The corrections (2), Fig.\ \ref{correl}b, can be absorbed in the nucleon
form factor (in the present approach, this form factor will appear 
after averaging over quark and antiquark coordinates inside a nucleon, 
cf.\ Section IV). Moreover, on the time scale of the nuclear collision gluon
exchange inside a nucleon can be neglected.
Similarly, we argue that corrections of type (3), Fig.\ 
\ref{correl}c, are nuclear structure effects and not important
on the scales of interest here. More rigorously, it was shown in \cite{yuri'}
that diagrams of such type are exponentially suppressed in the
covariant gauge. Corrections (1), Fig.\ \ref{correl}a, and (4),
Fig.\ \ref{correl}d, are higher order quantum effects to the
scattering process and suppressed by additional powers of $g$. 
Note that the corrections (2--4) induce {\em correlations\/} between
the two charges directly involved in the gluon emission process and other
charges.
Following the above arguments, these types of correlations will be neglected.

\begin{figure}
\begin{center}
\epsfxsize=6cm
\epsfysize=3cm
\leavevmode
\hbox{ \epsffile{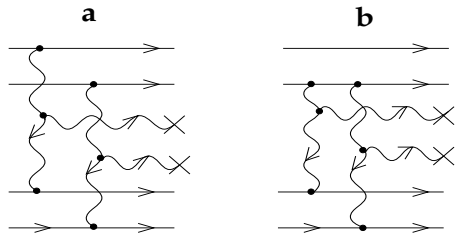}}
\end{center}
\caption{2--gluon emission processes, a) from collisions of different
charges, b) from collisions where one charge collides twice.}
\label{emiss}
\end{figure}

The amplitudes (\ref{ampl})
for states containing $n$ gluons are simply given by {\em products\/} 
of amplitudes for one--gluon states.
Each one--gluon amplitude contains $4\, A_1 A_2$ terms.
A product of $n$ such amplitudes, corresponding to emission of $n$ 
gluons, contains $(4\, A_1 A_2)^n$ terms in total. Of these,
$(2A_1)!\, (2A_2)!\, /\, (2A_1-n)!\,(2A_2-n)!$ terms 
correspond to processes where different
charges collide {\em once\/} (in a given order) and emit a gluon,
see e.g.\ Fig.\ {\ref{emiss}}a. These terms are identical to
those one would obtain in a perturbative calculation in terms of diagrams, 
since these diagrams contain no multiple scatterings of a single charge.

The remaining terms correspond to processes where a given charge 
collides and emits a gluon {\em more than once}, cf.\ Fig.\ {\ref{emiss}}b. 
In the coherent state, such a process is simply given by products
of one--gluon amplitudes. Before and after each one--gluon emission process
(as represented by the diagrams of Fig.\ \ref{diag}),
the charge is on-shell. On the other hand,
in the corresponding diagrammatic calculation the charge is
off-shell. In fact, since the nuclei are highly Lorentz--contracted 
and since the distance between subsequent collisions is inversely 
proportional to the corresponding Lorentz factor,
the charge is rather strongly off-shell. 
The coherent state approach as discussed here fails to account for this and
thus to give the correct description
for the emission of more than one gluon from the same quark line.

In other words, multiple collisions of the same charge induce
{\em correlations\/} between the $n$ emitted gluons. Such correlations
are correctly accounted for in a quantum calculation in terms of
diagrams. The coherent state approach, on the other hand, neglects
these correlations and assumes that {\em each\/} individual gluon
is emitted in an {\em independent\/} binary collision between
color charges. This is what gives rise to the Poisson distribution
(\ref{Poisson}) for the emission of $n$ gluons.
Note that in addition to these correlations from multiple
collisions, there are also higher order quantum corrections to the $n$--gluon
emission process, which introduce correlations. These correspond to
diagrams similar as in Fig.\ \ref{correl}.
According to the above discussion, these correlations will be neglected here.

\vspace*{2.4cm}
\begin{figure} \hspace*{2.5cm} 
\psfig{figure=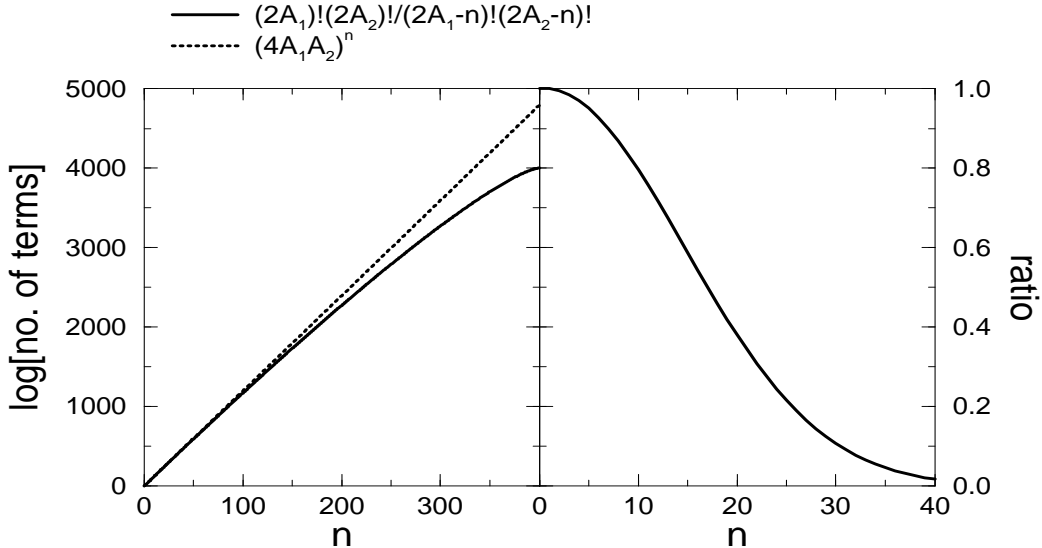,width=2.7in,height=3.5in,angle=-90}
\vspace*{-4cm}
\caption{Left panel: The logarithm of the number of terms corresponding to 
uncorrelated emission of $n$ gluons (solid) and the logarithm
of the total number of terms (dotted) vs.\ $n$. The right panel shows the
corresponding ratio of these numbers. For the sake of definiteness,
$A_1=A_2=200$ was assumed.}
\label{combinatorics}
\end{figure}

In the following we want to derive a sufficient condition under which
circumstances gluon emission from independent collisions of different
charges dominates that from multiple collisions of the same charge. We
argue that if this condition is fulfilled, the coherent state approach 
is applicable for the description of gluon radiation in ultrarelativistic 
nuclear collisions. We begin by noting that for small $n$ and large 
$A_1,\, A_2$, the number of terms corresponding to multiple collisions of
the same charge is negligible as compared to the number of terms
where charges collide only once and emit a gluon.
This is graphically shown in Fig.\ \ref{combinatorics} where we
plot the latter number (solid line) and the total number of terms
(dashed). The right-hand panel shows the ratio of these numbers.
Using Stirling's formula for large $A_1, A_2$, this ratio can be 
approximated by $\exp [-n(n-1)\,(A_1+A_2)/ 4\, A_1 A_2]$, i.e., an 
approximate Gaussian behavior with $n$, as indeed seen in Fig.\
\ref{combinatorics}. Obviously, for
$n(n-1) \ll 4\, A_1 A_2/(A_1+A_2)$ the number of terms where charges collide
only once and emit a gluon dominates the number of terms
where gluons are emitted in multiple collisions of the same charge.
In the situation of interest to us, $A \equiv A_1 \simeq A_2 \gg 1$, 
such that for the following discussion we may simplify this condition
as $n^2 \ll 2\, A$. This condition is a {\em sufficient\/} criterion to 
neglect gluon emission from multiple collisions of the same charge.

For $n^2 \ll 2\,A$, the $n$ gluons are more likely to be 
produced in independent collisions
of different charges than in multiple collisions of the same charge.
In this case, we expect the description in terms of a coherent state to be
(approximately) correct. For large $n^2 \sim 2\, A$, 
on the other hand, the emission of gluons from multiple collisions
of the same charge is not negligible. The emission process for these gluons 
is incorrectly treated in the coherent state approach, and the discrepancy
to the correct (quantum) result can in principle become large.
Note, however, that a quantitative calculation is required to prove
that this is indeed the case. It may be that even for large $n^2$
the deviation of the coherent state approach from the true answer
remains small. Therefore, the criterion $n^2 \ll 2\,A$ 
is rather conservative.

In order to minimize the deviation of the coherent state approach from
the correct result, we have to require that the {\em probability\/} of
occurrence of quantum states with a large number of gluons (which
are incorrectly treated in the coherent state approach) is small. 
Since that probability follows the Poisson distribution (\ref{Poisson})
in a coherent state, this happens exactly when the 
expectation value of the total gluon number in the coherent state,
$\bar{N}$, is small. Our criterion then becomes $\bar{N}^2 \ll 2\, A$. 
Note that if $\bar{N}^2 \ll 2\, A$ is fulfilled, it certainly holds for any
subset of the total number $\bar{N}$ of radiated gluons as well.
A subset of particular interest to us is, for instance,
the number of gluons per rapidity, $d \bar{N}/dy$. Moreover, even if
it does not hold for the total gluon number, $\bar{N}$, it may still
be satisfied for a smaller subset such as $d \bar{N}/dy$.
This is obvious, since gluons in that subset are more likely
to be less correlated.
In Section V we shall discuss this criterion in a more quantitative
way for the situation of an ultrarelativistic nuclear collision.

Note that this result is at first sight rather surprising. Naively, 
one would argue that the classical description becomes better when the 
number of emitted gluons is large, such that the notion of a
classical field becomes reasonable. 
As the description in terms of a coherent state is the 
closest analogue to the classical approach, one would expect a similar
condition to hold in this case, too, for instance, that the expectation
value of the number of gluons in the state $|\, \tilde{J}\, \rangle$
should be large. In contrast, from the preceding arguments one
concludes that the coherent state approach is more viable
for $\bar{N}^2 \ll 2\,A$.
This apparent contradiction is resolved noting that {\em not\/} a
large $\bar{N}$ is the essential criterion but rather
that the emission of the $n$ gluons in an $n$--gluon state happens 
in an {\em uncorrelated\/} fashion. This is exactly what gives rise
to the Poisson distribution (\ref{Poisson}) for the probability to 
find $n$ gluons in the coherent state. If there are correlations
in the emission of the gluons, the probability would no longer
be a Poisson distribution, and the coherent state approach is
no longer applicable.
As discussed above, multiple collisions of the same charge are the major 
source of correlations and thus for potential deviations from the 
coherent state approach. The condition $\bar{N}^2 \ll 2\,A$ ensures that 
such correlations are negligible.

\section{The event-averaged gluon number distribution}

The coherent state $| \tilde{J} \, \rangle$ depends on the 
initial color orientations of the quarks and antiquarks, 
$(T^a_i)\, ,\,\, (\tilde{T}^b_j)$, and
their coordinates $x_i,\, x_i',\, y_j,\, y_j'$ in a single
collision event. These coordinates and color orientations are
supposed to be fixed in the initial state. The 
coherent state is therefore a pure state. In other words, the 
phase information about the emitted gluons is complete, and
all gluons are emitted {\em coherently}. This is certainly unphysical,
because the emission of two gluons is {\em incoherent\/}
if the spatial distance between their emission points is larger than their
inverse momentum. In the classical approach of Refs.\
\cite{Alex,Larry,yuri,yuridirk} as well as in the coherent state
formulation discussed here, this incoherence is introduced
{\em a posteriori\/} by averaging over all possible color orientations 
and coordinates. In the following, we therefore compute the respective 
average of the expectation value of the gluon number distribution 
(\ref{dbarndk}). We shall refer to this average as {\em event average\/}.
In an extension of the treatment in \cite{yuridirk} 
we shall now also take a possible finite impact parameter $\underline{b}$ 
of the nuclear collision into account.

Similar as in \cite{yuridirk}, we introduce center-of-mass coordinates for
nucleon $i$, $(X_{-i}, {\underline X}_i)$, 
{\em as measured in the center of mass of nucleus 1}, and for nucleon $j$, 
$(Y_{+j},{\underline Y}_j)$, {\em as measured
in the center of mass of nucleus 2}, and relative coordinates 
$\Delta x_{-i}, \, \Delta {\underline x_i}, \, \Delta y_{+j},\, 
\Delta {\underline y}_j$ for the quarks and antiquarks inside each 
nucleon, as measured from the center of mass of the individual nucleon. 
Thus, if we assume that the center of mass of nucleus 1 moves along 
the $z$--axis with velocity $v_1=1$, and the center of mass of 
nucleus 2 parallel to the $z$--axis at a transverse distance 
(impact parameter) $\underline{b}$ and with velocity $v_2 = -1$, then
\begin{mathletters}
\begin{eqnarray}
x_{-i} = X_{-i} + \frac{\Delta x_{-i}}{2}\,\, , \,\,\, x_{-i}' =
X_{-i} - \frac{\Delta x_{-i}}{2} \,\,\, & , & \,\,\, y_{+j} = Y_{+j} +
\frac{\Delta y_{+j}}{2} \,\, , \,\,\, y_{+j}' = Y_{+j} - \frac{\Delta
y_{+j}}{2}\,\, , \\ {\underline x}_i = {\underline X}_i + \frac{\Delta
{\underline x}_i}{2} \,\, , \,\,\,{\underline x}_i' = {\underline X}_i
- \frac{\Delta {\underline x}_i}{2} \,\, & , & \,\,\, {\underline y}_j
= \underline{b}+ {\underline Y}_j + 
\frac{\Delta {\underline y}_j}{2} \,\,, \,\,\,
{\underline y}_j' = \underline{b} + {\underline Y}_j - \frac{\Delta 
{\underline y}_j}{2} \,\, .
\end{eqnarray}
\end{mathletters}
In another generalization of the treatment in \cite{yuridirk}
we do no longer consider nucleons and nuclei to be ``cylindrical''.
The event average is therefore defined as
\begin{eqnarray}
\langle\, \cdot \, \rangle & \equiv &
\prod_{i=1}^{A_1} \int_1 \frac{d^2 \underline{X}_i\, d X_{-i}}{4\pi R_1^3\, /
\, 3 \sqrt{2}\, \gamma}\, \int_i \frac{d^2 (\Delta \underline{x}_i/2) \,
d(\Delta x_{-i}/2)}{4 \pi a^3\, /\, 3 \sqrt{2}\, \gamma}
\nonumber \\
& \times & \prod_{j=1}^{A_2} \int_2 \frac{d^2 
\underline{Y}_j \,d Y_{+j} }{4\pi R_2^3\, /\, 3 \sqrt{2}\, \gamma}\,
\int_j \frac{d^2 (\Delta \underline{y}_j/2)\,d( \Delta y_{+j}/2) }{
4 \pi a^3\, /\, 3 \sqrt{2}\, \gamma}
\, \, \frac{1}{N_c^2}\,\, {\rm tr}\, \left[ \, 
\cdot \, \right] \,\,\, . \label{ensave}
\end{eqnarray}
Here, $R_1,\, R_2$ are the radii of nucleus 1 and 2, respectively, $a$
is the nucleon radius, $\gamma$ the Lorentz factor in
the center-of-mass frame of the nuclei, and the trace is taken over
the color space of nucleon $i$ and $j$. The 
integrations over ``$1$'' and ``$2$'' run over the volume of nucleus 1 and
nucleus 2, respectively. Integrations over ``$i$'' and ``$j$'' run
over the corresponding volumes of nucleon $i$ and $j$. All
volumes are Lorentz--contracted spheres.

We are now in the position to calculate the {\em event-averaged\/}
expectation value for the gluon number distribution, Eq.\
(\ref{dbarndk}). If we follow the same steps as in \cite{yuridirk} 
(and take ``cylindrical'' nucleons and nuclei) we indeed obtain the 
same answer, Eq.\ (39) of \cite{yuridirk}. One of the assumptions of 
that derivation was, however, that the nuclei are (infinitely) large, 
cf.\ Eq.\ (35) of \cite{yuridirk}. In that case, the notion of a finite 
impact parameter is obsolete.

In the following we outline the derivation of the gluon
number distribution in the case of a nuclear collision at
finite impact parameter $\underline{b}$ and for spherical nucleons
and nuclei. Inserting the current (\ref{current03k}) into Eq.\ 
(\ref{dbarndk}), the expectation value of the gluon number distribution 
in the coherent state $|\, \tilde{J}\, \rangle$ becomes
\begin{eqnarray}\label{expNdist}
\frac{d\bar{N}}{dy\, d^2 \underline{k}} & = & 4\, \frac{g^6}{2(2\pi)^7} \,
f^{abc}\, f^{ade}  \,\, \frac{1}{{\underline k}^2}   
\int d^2 {\underline q}_1 \, d^2 {\underline q}_2 \,
\frac{{\underline q}_1 \cdot {\underline q}_2 \, {\underline k}^2 + 
{\underline q}_1^2\, {\underline q}_2^2 - {\underline q}_1^2\, 
{\underline k} \cdot {\underline q}_2 -{\underline q}_2^2 \, 
{\underline k} \cdot {\underline q}_1}{ {\underline q}_1^2\,\, 
{\underline q}_2^2 \,\, (\underline{k} - \underline{q}_1)^2\,\, 
(\underline{k}-\underline{q}_2)^2} \,\, \\
& \times & \sum_{i,k=1}^{A_1} \sum_{j,l=1}^{A_2} 
(T_i^b) \, (\tilde{T}_j^c) \, (T_k^d) \, (\tilde{T}_l^e) \,\, 
 {\cal P}(k_+,\underline{q}_1;x_i) \, 
 {\cal P}(k_-,\underline{k} - \underline{q}_1;y_j)\,
 {\cal P}^*(k_+, \underline{q}_2;x_k)\, 
 {\cal P}^*(k_-, \underline{k}- \underline{q}_2;y_l) \,\, ,
 \nonumber
\end{eqnarray}
where $k^\mu$ is on-shell and
\begin{equation}
{\cal P}(k_+,\underline{q};x_i) \,\, \equiv \,\, e^{ i k_+ x_{-i} - i
{\underline q} \cdot {\underline x}_i }\,\, - \,\, e^{ i k_+ x_{-i}' -
i {\underline q} \cdot {\underline x}_i'} \,\, .
\end{equation}
We now perform the event average (\ref{ensave}). The averaging
over color is done utilizing ${\rm tr}\,[\, (T_i^b) \,(\tilde{T}_j^c)\,
(T_k^d)\, (\tilde{T}_l^e)\,] = \delta_{ik} \, \delta^{bd} \delta_{jl}
\,\delta^{ce} /4$, and $f^{abc} \, f^{abc} = N_c (N_c^2-1)$, with the
result
\begin{equation} \label{dndyd2k}
\frac{d \langle\, \bar{N}\, \rangle}{dy \, d^2 \underline{k}} = 
\frac{g^6}{2 (2\pi)^7}\, \frac{N_c^2-1}{N_c} \,
 \frac{1}{{\underline k}^2} \int
d^2 {\underline q}_1 \, d^2 {\underline q}_2 \, 
\frac{{\underline q}_1 \cdot {\underline q}_2 \, {\underline k}^2
 + {\underline q}_1^2\, {\underline q}_2^2 - {\underline q}_1^2\, 
{\underline k} \cdot {\underline q}_2
-{\underline q}_2^2 \, {\underline k} \cdot {\underline q}_1}{ 
{\underline q}_1^2\,\,  {\underline q}_2^2 
\,\, ({\underline k - q}_1)^2\,\, ({\underline k-q}_2)^2}\,\, 
\langle \, \tilde{\cal P} (k,\underline{q}_1, \underline{q}_2)\, 
\rangle \, ,
\end{equation}
where 
\begin{eqnarray}
\tilde{\cal P}(k,\underline{q}_1, \underline{q}_2) & = &
\sum_{i=1}^{A_1} e^{-i ({\underline q}_1 - {\underline q}_2) \cdot
{\underline X}_i } \left[ \,\, e^{-i ({\underline q}_1 - {\underline
q}_2) \cdot \Delta {\underline x}_i/2} - e^{ik_+ \Delta x_{-i}
-i({\underline q}_1 + {\underline q}_2) \cdot \Delta {\underline
x}_i/2} + {\rm c.c.}\,\, \right] \nonumber \\ & \times &
\sum_{j=1}^{A_2} e^{i ({\underline q}_1 - {\underline q}_2) \cdot
(\underline{b} + {\underline Y}_j) } 
\left[ \,\, e^{i ({\underline q}_1 - {\underline
q}_2) \cdot \Delta {\underline y}_j/2} - e^{ik_- \Delta y_{+j} -i
{\underline k} \cdot \Delta {\underline y}_j + i({\underline q}_1 +
{\underline q}_2) \cdot \Delta {\underline y}_j/2} + {\rm c.c.}\,\,
\right] \,\, .
\end{eqnarray}
We now average over the light-cone variables
$\Delta x_{-i}$. The corresponding integration
runs between $\pm f/\gamma \equiv \pm \sqrt{a^2 - (\Delta 
\underline{x}_i/2)^2}/\sqrt{2}\gamma$ and gives
\begin{equation} \label{longave}
\int^{f/\gamma}_{-f/\gamma} 
d\left( \frac{ \Delta x_{-i}}{2}\right) \, e^{ik_+ \Delta x_{-i}} =
\frac{\sin[ 2\, k_+ f/ \gamma]}{ k_+ } 
\rightarrow \frac{2\, f}{\gamma} \,\,\,\,\,\,\, 
(\gamma \rightarrow \infty) \,\, .
\end{equation}
With an analogous relation for the average over $\Delta y_{+j}$, the
longitudinal momentum dependence in the phase factor drops out. 
The average over the light-cone variables $X_{-i}, \, Y_{+j}$ is
even simpler, since $\tilde{\cal P}$ does not depend on them. The 
integration, however, yields a factor analogous to (\ref{longave}).
For the average over transverse coordinates we then use \cite{GR}
\begin{equation}
\frac{1}{2\pi\, R^2/3} 
\int_{|\underline{x}| \leq R} d^2 \underline{x} \,\, e^{i \underline{q}
\cdot \underline{x}}\,\, \sqrt{1 - \frac{\underline{x}^2}{R^2}}
= \frac{3}{\underline{q}^2 R^2} \left[ \frac{ \sin (|\underline{q}|R)}{
|\underline{q}|R} - \cos (|\underline{q}|R) \right]
= \frac{3\, j_1(|\underline{q}|R)}{|\underline{q}|R} 
\equiv \Delta(|\underline{q}|R)\,\, ,
\end{equation}
where $j_1$ is a spherical Bessel function
[note that for ``cylindrical'' nucleons and nuclei, $\Delta(x)
\equiv 2\, J_1(x)/x$ \cite{yuridirk}],
with the result
\begin{eqnarray} \label{tilPav}
\langle\, \tilde{\cal P} (k,\underline{q}_1, \underline{q}_2)\,
\rangle & = & 4\, A_1 A_2\,\, e^{i (\underline{q}_1 - \underline{q}_2)
\cdot \underline{b}} \,\, \Delta(|\underline{q}_1 - \underline{q}_2| R_1) \,\,
\Delta (|\underline{q}_1 - \underline{q}_2| R_2) \\ 
& \times & \left[ \Delta(|\underline{q}_1 - \underline{q}_2| a)
- \Delta (|\underline{q}_1 + \underline{q}_2| a) \right]
\, \left[\Delta(|\underline{q}_1 -\underline{q}_2| a)
- \Delta(|2 \underline{k}- \underline{q}_1 -
\underline{q}_2| a) \right] \,\, .\nonumber
\end{eqnarray}
$\Delta(x)$ is a rapidly decreasing function of its argument, thus
the factors $\Delta(|\underline{q}_1 - \underline{q}_2| R_{i}),
\, i=1,2,$ limit the difference
$|\underline{q}_1 - \underline{q}_2|$ to a scale of order 
${\rm min}\{\pi/R_1,\pi/R_2\}$. 
That scale is much smaller than the typical scale of
variation $\sim 1/a$ of the remaining terms in (\ref{tilPav}) and in
the integral over $\underline{q}_1,\, \underline{q}_2$ in Eq.\ 
(\ref{dndyd2k}). [An exception is the factor
$e^{i(\underline{q}_1 - \underline{q}_2)\cdot \underline{b}}$
which, for large impact parameters $|\underline{b}| \sim R_i,\, i=1,2,$
then varies over the same small scale $\pi/R_i, \, i=1,2$, and consequently
must not be simply approximated by 1.]
In these terms, we may therefore take to good approximation
$\underline{q}_2 \simeq \underline{q}_1$. [This equality was achieved 
by quite similar means in \cite{yuridirk} via the assumption that if 
$R_1 > R_2 \gg a$, then $\Delta(|\underline{q}_1 - \underline{q}_2| 
R_1) \simeq 4\pi/R_1^2 \, \delta(\underline{q}_1 - \underline{q}_2)$,
cf.\ Eq.\ (35) of \cite{yuridirk}.]

The final result for the event-averaged gluon number distribution reads
\begin{equation} \label{dndyd2kfinal}
\frac{d \langle \, \bar{N}\, \rangle}{ dy\, d^2 {\underline k}} = \pi\,\,
\frac{g^6}{(2\pi)^6}\, \frac{N_c^2 -1}{N_c}\, \, \langle \,
T_{A_1A_2}(\underline{b})\,\rangle \,\,
\frac{4}{{\underline k}^2}\,\, \int d^2
{\underline q} \,\, \frac{F(|{\underline q}|a)}{{\underline q}^2} \,\,
\frac{F(|\underline{k}-\underline{q}|a)}{(\underline{k}-
\underline{q})^2} \,\, ,
\end{equation}
with $F(x) = 1 - \Delta(2x)$, and where we defined
\begin{equation} \label{TAA}
\langle \, T_{A_1A_2}(\underline{b}) \,
\rangle \equiv \int \frac{d^2 \underline{l}}{(2 \pi)^2}
\,\, e^{i \underline{l} \cdot \underline{b}} \,\,
A_1 A_2 \,\, \Delta(|\underline{l}| R_1)\,
\Delta(|\underline{l}| R_2) \,\,\, .
\end{equation}
While $\Delta$  in Eq.\ (\ref{dndyd2kfinal}) (cf.\ the definition of $F$)
stands for the {\em nucleonic\/} form factor \cite{gunion}, in 
(\ref{TAA}) it represents the {\em nuclear\/} form factor.

We will now show that $\langle \, T_{A_1A_2}(\underline{b}) \,
\rangle $ is the {\em event-averaged number of binary nucleon--nucleon
collisions per transverse area\/} in an $A_1+A_2$--collision at 
impact parameter $\underline{b}$ (the so-called nuclear {\em thickness\/}
function). Note that for a central collision, uniform nuclear
density distributions, and ``cylindrical'' nuclei, this 
factor is $A_1A_2/\pi R_1^2$ (for nucleus 1 being the larger of the two 
nuclei), and we thus re-obtain the answer Eqs.\ (38,39) of \cite{yuridirk}
[up to the modified form factor $F(x)$].

To prove this assertion, we revert the
averaging over the center-of-mass coordinates of the nucleons, i.e.,
\begin{equation}
 T_{A_1A_2}(\underline{b}) = \int \frac{d^2 \underline{l}}{(2 \pi)^2}
\,\, \sum_{i,j} e^{-i \underline{l} \cdot (\underline{X}_i - 
\underline{Y}_j -\underline{b})}\,\, ,
\end{equation}
and introduce a factor of 
\begin{equation}
1 \equiv \int d^2 \underline{X}\, d^2 \underline{Y}\, 
\delta(\underline{X}-\underline{X}_i)\, 
\delta(\underline{Y}-\underline{Y}_j)
\end{equation}
under the sum. This enables us to perform the $\underline{l}$--integration.
Introducing the baryon number distributions for nucleus 1 and 2,
\begin{equation}
\rho_{A_1}(\underline{X},Z_1) = \sum_{i=1}^{A_1} \, 
\delta (\underline{X}-\underline{X}_i)\,\delta(Z_1-Z_{i})\,\,\, , \,\,\,\, 
\rho_{A_2}(\underline{Y},Z_2) = \sum_{j=1}^{A_2} \, 
\delta (\underline{Y}-\underline{Y}_j)\,\delta(Z_2-Z_{j})\,\,\, , 
\end{equation}
we realize that
\begin{equation}
 T_{A_1A_2}(\underline{b}) = \int d^2 \underline{X}\, 
\int  d Z_1 \, \rho_{A_1}(\underline{X},Z_1)\,
\int d Z_2\, \rho_{A_2} (\underline{X}-\underline{b},Z_2)
\end{equation}
is {\em the number of binary nucleon--nucleon collisions per
transverse area\/} for an $A_1+A_2$--collision 
at impact parameter $\underline{b}$.
The event average of $T_{A_1A_2}(\underline{b})$ is simply
\begin{equation}
 \langle \, T_{A_1A_2}(\underline{b})\, \rangle
 = \int d^2 \underline{X} \,   \int  
d Z_1 \, \langle \, \rho_{A_1}(\underline{X},Z_1)\,\rangle \,
\int d Z_2\, \langle \, \rho_{A_2} (\underline{X}-\underline{b},Z_2)\, 
\rangle \,\,\, .
\end{equation}
For Gaussian nuclear density distributions, $\rho_A ({\bf x}) = A\,
\exp [-{\bf x}^2/R^2 ]\, (\pi R^2)^{-3/2}$, we obtain the well-known result
$ \langle \, T_{A_1A_2}(\underline{b})\, \rangle = A_1A_2 \, \exp
[-\underline{b}^2/(R_1^2+R_2^2)] / \pi (R_1^2 +R_2^2)$.

We have thus generalized the result of \cite{yuridirk} for
the gluon number distribution to collisions at finite impact parameter.
As expected, the gluon number decreases with the impact parameter
according to the decrease of $ \langle\, T_{A_1A_2}(\underline{b})\, \rangle$
with $\underline{b}$. We also conclude from the results of this section
that the classical solution \cite{yuridirk} and the coherent state 
approach yield the same expression for the gluon number distribution.

\section{Conclusions}

In this work we have exhibited the relationship between classical 
gluon radiation in ultrarelativistic nuclear collisions as calculated in
\cite{Alex,MGLMcL,yuridirk} and corresponding quantum radiation 
in a coherent state.
We have demonstrated that the coherent state formalism yields the
same result for the gluon number distribution as the classical approach.
A condition for the coherent state approach to represent
a valid description of gluon radiation in ultrarelativistic nuclear
collisions was found to be that the expectation value of 
the gluon number in the coherent state has to be small compared
to the number of sources in the (classical) source current, since then
the emission of individual gluons happens in an
uncorrelated fashion. The event-averaged version of this criterion
reads $\langle \, \bar{N}\, \rangle^2 \ll Z_{\rm eff}$, where
$Z_{\rm eff}$ is the effective number of sources in the classical
source current, which is, for the simple nuclear model considered
here, equal to $2\, A$.

This condition is similar to the criterion 
$\mu^2 \gg \underline{k}^2 \gg \Lambda_{\rm QCD}^2$ for the applicability
of the classical approach of Ref.\ \cite{Larry}, since the latter states
that the (transverse) area density of
source charges should be large on transverse momentum scales of interest
(such that the source can be described classically) while the 
number of field quanta $\sim \mu^2/\underline{k}^2$ associated with 
these sources must not be too large, i.e., it must be small as
compared to $\mu^2/\Lambda_{\rm QCD}^2$.
The problem is, however, that if one considers
valence charges only, such as in the approach advertised here and
in Refs.\ \cite{yuri,yuridirk}, in the case of interest, i.e., for 
instance for a $Au+Au$--collision at RHIC energies, $\sqrt{s} = 200$ AGeV, 
$\mu^2 = (N_c^2-1) A/ (2N_c^2\, \pi R^2)$ \cite{yuri} is only
about $(160\, {\rm MeV})^2$, and thus of the order of $\Lambda_{\rm QCD}^2$
rather than much larger. It was therefore argued \cite{MGLMcL,Jamal}
that besides the valence charges considered here, one
should also add quark and gluon charges arising from the nucleonic
parton sea.

\begin{figure} \hspace*{2.5cm} 
\psfig{figure=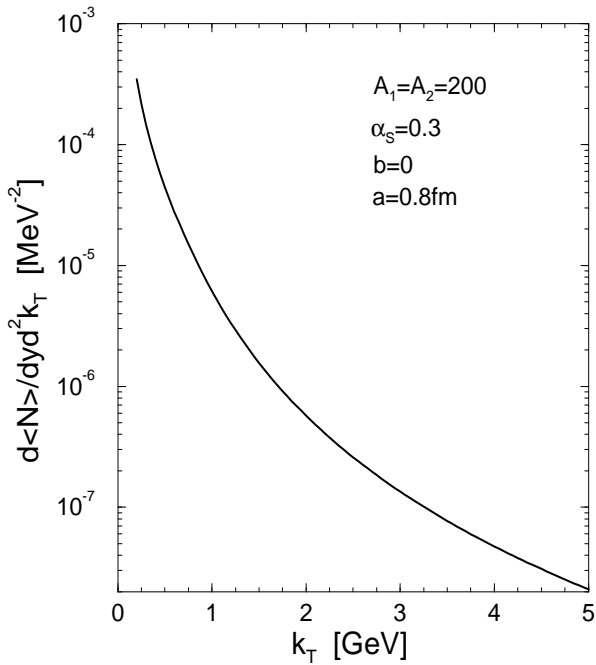,width=2.7in,height=2.5in,angle=-90}
\caption{The gluon number distribution for a central $A_1+A_2$--collision.}
\label{gluespec}
\end{figure}

On the other hand, in order to estimate whether the condition
$\langle \,\bar{N}\, \rangle^2 \ll 2\, A$ 
found here in the framework of the coherent state approach is quantitatively 
fulfilled, we compute the event-averaged gluon number distribution from Eq.\
(\ref{dndyd2kfinal}). [We use the form factor for ``cylindrical''
nucleons, since in that case an analytical expression for
the gluon number distribution is known \cite{yuridirk}; the difference
between the form factors in the cylindrical and spherical
cases is small.] The result is shown in Fig.\ \ref{gluespec} for
a central collision with $A_1=A_2=200$. The strong coupling constant
was taken as $\alpha_S \equiv g^2/4 \pi = 0.3$, the nucleon radius
(entering the nucleonic form factors) was assumed to be $a=0.8$ fm.

\vspace*{2.5cm}

As noted in \cite{yuridirk}, the infrared behavior of the distribution is
$\sim 1/\underline{k}^2$, due to the nucleonic form factors,
not $\sim \log (\underline{k}^2)/\underline{k}^4$ as claimed
in \cite{Alex,MGLMcL}. Therefore, $d\langle\, \bar{N}\, \rangle /dy$
diverges only logarithmically. This divergence is the well-known
infrared catastrophe \cite{itzykson}. Physically, it will
be regulated on a typical hadronic mass scale due to confinement. 
Here we simply introduce a lower transverse momentum cut-off 
$|\underline{k}_{\rm min}|$ to estimate $d\langle\, \bar{N}\, \rangle /dy$. 
Let us first choose the rather low $|\underline{k}_{\rm min}| = 
\Lambda_{\rm QCD} \simeq 200$ MeV. From a numerical integration of
the spectrum in Fig.\ \ref{gluespec} up to $|\underline{k}| = 10.2$
GeV we obtain for the gluon rapidity density 
$d\langle\, \bar{N}\, \rangle /dy \simeq 140$.
The rapidity distribution is constant in $y$, since the radiating
charges are assumed to move along the light-cone.  However,
since the classical approach fails anyway in the fragmentation 
regions, $y \sim \pm y_{\rm CM}$ \cite{gunion}, one should not
consider gluons other than such produced around midrapidity 
$y \simeq 0$. We shall therefore use the computed $d \langle \, \bar{N}
\, \rangle /dy$ (at midrapidity) to check
the condition for the viability of the coherent state approach.
As discussed in Section III, it is always possible to use 
a subset for this purpose rather than the total number of gluons.
We thus find that $(d\langle \, \bar{N} \, \rangle/dy)^2 \simeq 20000 >
2\, A = 400$, i.e., the criterion found in Section III is not 
satisfied. In other words, the resulting gluon multiplicity is so large
that a part of the gluons must have been produced in multiple collisions 
of the same charge.

On the other hand, gluons with $|\underline{k}| \sim \Lambda_{\rm QCD}$ can
certainly no longer be described in perturbative terms. If we consider an
even smaller subset and take
an infrared cut-off of $1$ GeV, we obtain 
$d\langle\, \bar{N}\, \rangle /dy \simeq 26$. For an even larger cut-off of
$2$ GeV \cite{HIJING}, $d\langle\, \bar{N}\, \rangle /dy \simeq 8$.
For this cut-off, the gluon multiplicity is small enough that these gluons
are likely to be produced in a statistically independent way.
Thus, the coherent state approach to describe the 
production of such gluons seems justified.

This result establishes that, although $\mu^2 \sim \Lambda_{\rm QCD}^2$,
i.e., in principle violation of the fundamental assumption of the 
approach of Ref.\ \cite{Larry}, for practical purposes, i.e.,
for midrapidity gluons with perturbative transverse momenta 
$|\underline{k}| > 1-2$ GeV and produced in collisions between valence
charges, a description in terms of a coherent state generated by a 
classical source of color charges is likely to be reasonable. 
The criterion $\mu^2 \gg \underline{k}^2 \gg \Lambda_{\rm QCD}^2$ 
is certainly a sufficient condition for the applicability of 
the classical approach. Here we have established
that it might actually only be necessary to fulfill
$\langle\, \bar{N}\, \rangle^2 \ll Z_{\rm eff}$, with
$Z_{\rm eff}$ being the effective number of classical sources
(and the total number of gluons squared, $\langle \, \bar{N} \, \rangle^2$, 
possibly replaced by a smaller subset of this number).
If a quantum calculation of gluon emission from multiple
collisions of the same valence charge proves that the deviation from
the coherent state result for this process is small, then the
range of applicability of the coherent state approach is, for
practical purposes, even larger. Note that this is the case also for 
a smaller coupling constant.

A gluon rapidity density of $d\langle\, \bar{N}\, \rangle /dy \sim 140$ 
or even less for a central, ultrarelativistic
collision of $A=200$--nuclei appears rather small.
The reason is that, as already mentioned above, for $A=200$ the average 
transverse color charge density squared
$\mu^2 = (N_c^2-1) A/ (2N_c^2\, \pi R^2)$ \cite{yuri} is only
about $(160\, {\rm MeV})^2$. If one adds the charge contained in the
nucleonic parton sea \cite{MGLMcL,Jamal}, then
$\mu \sim 400$ MeV for collisions at RHIC energies \cite{MGLMcL}, which 
increases the gluon rapidity density by a factor of 40 (for fixed
nuclear radius $R$). For $|\underline{k}_{\rm min}| = 1$ GeV, this leads to
$d\langle\, \bar{N}\, \rangle /dy \sim 1000$, for 
$|\underline{k}_{\rm min}| = 2$ GeV, we obtain
$d\langle\, \bar{N}\, \rangle /dy \sim 325$. These values are well
within in the established range \cite{klaus,HIJING} of several hundred 
gluons per unit rapidity, confirming the conclusion of Ref.\
\cite{MGLMcL} that the classical description of gluon radiation
and the conventional pQCD mini-jet approach give 
approximately the same results. 

One has to note, however, that increasing the effective
charge of the classical current by a factor $\beta$ (when adding the
partonic sea) increases the
gluon multiplicity by a factor $\beta^2$ (for fixed radii of the colliding
nuclei). Therefore, the criterion $(d\langle \,\bar{N}\, \rangle /dy)^2 \ll 
Z_{\rm eff}$, where now $Z_{\rm eff}$ is the effective valence 
{\em plus\/} sea charge, may be violated in the 
process of adding charges from the nucleonic parton sea.
In this case, multiple collisions of the same (valence or sea) charge
become again important, and a more detailed quantum 
calculation of these processes is mandatory \cite{sergei}.

Finally, we have generalized the result for the gluon number 
distribution found in \cite{yuridirk} to nuclear collisions at 
finite impact parameter. As expected \cite{MGLMcL}, 
the number of binary nucleon--nucleon collisions per transverse area
for given impact parameter, $T_{A_1A_2}(\underline{b})$, appears as
a prefactor in the final result.

In future applications it is important to incorporate non-Abelian
effects on the dynamical evolution of the radiated gluons to
study screening, damping, and, eventually, thermalization which
leads to the formation of 
the proposed quark--gluon--plasma state of nuclear matter.

\section*{Acknowledgments}

D.H.R.\ thanks Miklos Gyulassy and Yuri V.\ Kovchegov for valuable
discussions. The authors are indebted to Yuri V.\ Kovchegov 
for a critical reading of the manuscript.
This work was supported in part by the U.S.\
Department of Energy under Contract No.\ DE-FG02-96ER-40945.

\end{document}